\documentclass[11pt]{article}
\usepackage{amssymb,amsmath,bm,a4,amsthm,array,xy,longtable,graphicx}

\newcommand{\eeq}{\end{equation}}
\newcommand{\beq}{\begin{equation}}
\newcommand{\ba}{\begin{array}}
\newcommand{\ea}{\end{array}}
\newcommand{\bea}{\begin{eqnarray}}
\newcommand{\eea}{\end{eqnarray}}
\newcommand{\vev}[1]{\langle #1\rangle}

\newcommand{\preprintno}[1]{\vspace{-2cm}{\normalsize\begin{flushright}#1\end{flushright}}\vspace{1cm}}
\newcommand{\heth}{${^3}$He}
\newcommand{\hefo}{${^4}$He}
\newcommand{\lisi}{${^6}$Li}
\newcommand{\lise}{${^7}$Li}
\newcommand{\bese}{${^7}$Be}
\newcommand{\beei}{${^8}$Be}

\title{\preprintno{HD-THEP-07-10}Primordial nucleosynthesis as a probe of fundamental physics parameters}

\author{Thomas Dent,
	Steffen Stern and 
	Christof Wetterich \vspace{0.2cm}\\ 
	{\em Theoretische Physik, 16 Philosophenweg, Heidelberg 69120 GERMANY}
	}

\date{\today}

\begin{document}

\maketitle

\begin{abstract}\noindent
We analyze the effect of variation of fundamental couplings and mass scales on primordial nucleosynthesis in a systematic way. The first step establishes the response of primordial element abundances to the variation of a large number of nuclear physics parameters, including nuclear binding energies. We find a strong influence of the $n-p$ mass difference (for the \hefo\ abundance), of the nucleon mass (for deuterium) and of $A=3,4,7$ binding energies (for \heth, \lisi\ and \lise). A second step relates the nuclear parameters to the parameters of the Standard Model of particle physics. The deuterium, and, above all, \lise\ abundances depend strongly on the average light quark mass $\hat{m}\equiv (m_u+m_d)/2$. We calculate the behaviour of abundances when variations of fundamental parameters obey relations arising from grand unification. We also discuss the possibility of a substantial shift in the lithium abundance while the deuterium and \hefo\ abundances are only weakly affected.
\end{abstract}

\section{Introduction}
The constancy over space and time of the coupling strengths and particle masses in the Standard Model of particle physics is an assumption that should be tested.\footnote{For a general review of ``varying constants'' see \cite{Uzanreview}.} If a variation existed, it would violate the condition of Local Position Invariance contained in the Einstein equivalence principle of General Relativity: non-gravitational experiments measuring the same quantities at different times would give different results. Within a relativistically covariant setting, a variation arises due to the coupling of Standard Model particles to a scalar field whose cosmological value depends on time \cite{Jordan?,Bekenstein82}.

Dynamical dark energy or quintessence \cite{Wetterich88,RatraPeebles} is precisely due to a scalar field whose value continues to vary in recent cosmological epochs. Thus the possibility of time-varying couplings has been discussed \cite{Wetterich88_2} since the first suggestions of dynamical dark energy. Related ideas concern the dilaton or time-varying moduli fields in string theory \cite{DamourVeneziano}. Many dark energy candidates in higher-dimensional or string theories can be excluded because the time-variation of couplings is too strong. On the other hand, a detection of time-varying couplings would be a strong hint in favour of a dynamical dark energy, explaining the renewed interest in this subject \cite{DvaliZ,OliveP0104,WetterichProbing,WetterichCrossover,Parkinson04,Doran04}.

Such investigations have recently been further motivated by possible signals of a nonzero variation at redshifts about $0.5$--$4$, due to astrophysical absorption spectra which deviate from those found in the laboratory. By choosing the transitions to be measured, one can achieve sensitivity at the $10^{-5}$ level to the fine structure constant $\alpha$, the proton-electron mass ratio $\mu\equiv m_p/m_e$ \cite{Reinhold06}, the proton gyromagnetic ratio $g_p$, and various combinations of these \cite{Tzanavaris06}. The current observational situation is contradictory, with both positive \cite{Murphy03} and null \cite{Chand04,Levshakov} results, and continuing discussions of methods and of statistical and systematic errors \cite{Murphycrit}.

Other limits on the values of particle couplings and masses at (relatively) recent epochs arise from nuclear physics effects in the Oklo natural reactor at $z\simeq 0.2$ \cite{Gould07,Petrov05} and from long-lived $\beta$ decay isotopes in meteorites \cite{OliveRe}. Direct comparisons of atomic clocks at periods of a few years in the laboratory have also led to strong bounds at the level of $10^{-15}$ per year fractional variation \cite{atomic}. Since these are Earth- or Solar System-based probes their results may not be directly comparable with those from cosmologically distant absorption systems, without further theoretical assumptions. Clearly it is desirable to have as many bounds as possible at different redshifts and in different environments to probe possible variations in a model-independent way.

In this paper we consider possible variations at a much higher redshift, that of Big Bang nucleosynthesis (BBN) at $z \simeq 10^{10}$, which is currently the earliest time at which theories of nuclear and particle physics can be compared to cosmological observation in a controllable way \cite{PDGFields,Steigman05,Sarkar95}. It is remarkable that primordial abundances are influenced by every known force of interaction: gravity, through the expansion rate of the Universe; weak interactions, through neutron decay; electromagnetism, through the $n-p$ mass difference, nuclear binding energies, and Coulomb-suppressed and radiative nuclear reactions; and the strong interaction, throughout nuclear physics. 

Many studies of the effects on BBN of varying one or more parameters in particle physics have been performed in recent years \cite{CocNunes06,LiChu05,Chamoun05,Landau04,MSW,Dmitriev03,ScherrerGN,KnellerLambda,
YooScherrer,NollettLopez,Ichikawa02,Ivanchik01,DentF,BergstromIR,CampbellPrimordialSoup}; see also \cite{Uzanreview} and references therein. However, the use of BBN faces two major theoretical challenges. First, the degeneracy between many different variable parameters, to be compared with the small number of observable abundances. Only \hefo, \heth, deuterium and \lise\ have currently been measured to a level of accuracy which gives some prospect of comparison with theory, and of these \heth\ is subject to very large uncertainties in extrapolating back to the primordial abundance. We also consider \lisi, since it has been suggested that the stellar isotope has been measured \cite{Asplund05} and its primordial abundance can be predicted, although at such a low level that the primordial value may very easily be swamped by \lisi\ generation through astrophysical events or energetic late particle decay.

Secondly, theoretical uncertainty concerning the way in which the QCD parameters, in particular quark masses, affect nuclear forces, and thus nuclear binding energies and cross-sections. A derivation of nuclear forces from first principles, apart from the long-range attraction attributable to pion exchange, is lacking.\footnote{Recent efforts in lattice QCD \cite{Aokinuclear} have been encouraging.} Instead, various types of effective theories have been used in order to fit measured nuclear properties. These can yield a correct dependence on QCD parameters only to the extent that they reflect the underlying physics. There are also milder uncertainties in the dependence of certain nuclear reactions on $\alpha$, but in general the r{\^ o}le of electromagnetism is well understood. 

In this work we present a unified and systematic approach to these challenges, in two respects. First, we consider each variation in particle physics or ``fundamental'' parameters independently. Thus at linear order we can allow for {\em any}\/ theoretical scenario in which the variations of these parameters are subject to some kind of unified relation. Second, we identify which nuclear properties and reactions have significant influence on the variation of the primordial abundances. This is done by varying every relevant nuclear binding energy and cross-section in a reaction integration code, based on the Wagoner/Kawano code \cite{Wagoner69, Wagoner72,Kawano88,Kawano92} with updated numerical techniques, and noting the deviation of output abundances away from the unperturbed values. (Changes to the code are discussed in more detail in Appendix A.) This response to the nuclear parameters has then to be related to the variation of fundamental parameters via nuclear theory.

The results indicate where the sources of most theoretical uncertainty currently are, and suggest where improvements in nuclear theory are most needed. Our approach allows us to disentangle these problems. Our study of the dependence on nuclear parameters can be used as a basis, even if new theoretical insights modify the relations between nuclear physics and particle physics parameters used in this paper. The connection between the different levels of physical understanding arises by simple matrix multiplication of ``response matrices'' in a linear theory.

\section{Method}
We consider the set of primordial abundances $Y_a$ with $a=($D, \heth, \hefo, \lisi, \lise) and study its dependence on the variation of a set of nuclear physics parameters $X_i$. Here the index $i$ denotes different particle masses such as $m_p$ or $m_e$ and nuclear binding energies, as well as the neutron lifetime $\tau_n$ and couplings and mass scales such as the fine structure constant and Newton constant $G_{\rm N}$ that enter the calculation of nuclear abundances. Our central quantity is the response matrix $C$ with matrix elements \cite{MSW}
\beq \label{Cdef}
 c_{ai} = \frac{\partial \ln Y_a}{\partial \ln X_i}.
\eeq
It indicates the leading linear dependence for small deviations of the abundances about the values obtained given the nuclear parameters inferred from present laboratory experiments. The matrix $C$ is extracted by varying the quantities $X_i$ independently in the BBN code, a procedure which includes variation of the reaction cross-sections and rates that have a physical dependence on $X_i$. All variations in parameters are taken to be small, such that all necessary information can indeed be extracted from the response matrix. 

The second step involves the relation between a set of Standard Model parameters $G_k$ and the nuclear physics parameters $X_i$ encoded in a second response matrix $F$ with entries
\beq \label{Fdef}
 f_{ik} = \frac{\partial \ln X_i}{\partial \ln G_k}.
\eeq
This second step requires theoretical assumptions and contains, at present, substantial uncertainties, which we discuss in Section~\ref{fundamentals}.

The variation of abundances with respect to the fundamental parameters $G_k$ is expressed by the ``fundamental response matrix'' $R$ with elements $r_{ak}$:
\beq
 \frac{\Delta Y_a}{Y_a} = r_{ak} \frac{\Delta G_k}{G_k},
\eeq
The matrix $R$ is obtained from $C$ and $F$ by simple matrix multiplication:
\beq \label{RequalsCF}
R = CF.
\eeq

\subsection{Variation of dimensionless and dimensionful parameters}
It is sometimes overlooked that the variation of a dimensionful quantity is not physically well-defined: actually, one dimensionful quantity can only be measured by comparison with another. The fractional variations of dimensionful quantities are thus completely dependent on the choice of units; only dimensionless ratios are measurable.

In practice, one may choose a reference frame where some given mass scale is kept fixed. Popular frames are the Einstein frame where the Planck mass is kept fixed, or the Jordan frame where some particular particle physics scale is kept fixed. One may change from one frame to another by a Weyl transformation of the metric --- the time variation of dimensionless couplings and mass ratios is independent of the frame \cite{Wetterich88,Wetterich88_2}.\footnote{This supposes that the time interval is appropriately rescaled. However this is not an issue for BBN, since we do not consider the time derivative but rather the absolute variation between BBN and the present time.} In this paper we use a frame where the QCD invariant scale $\Lambda_c$ (sometimes written as $\Lambda_{\rm QCD}$) is kept constant. This is convenient for dealing with nuclear reactions, where the masses and energy scales are mainly determined by the strong interactions. Thus the variations of dimensionful parameters include implicitly some power of $\Lambda_c$. For example, if we take the electron mass $m_e$ as a parameter we are implicitly considering a variation of $m_e/\Lambda_c$. The implications of varying dimensionful parameters in QCD and nuclear physics are discussed further in Appendix~\ref{nucscaling}.

\section{Nuclear response matrix}
We first establish the matrix $C$ by a systematic variation of parameters in the BBN code. For this purpose we have modified the code as described in Appendix A, which also contains the details of our treatment of reaction rates.

\subsection{Nuclear parameters and response matrix}\label{nucparams}

We vary with respect to the following thirteen ``nuclear'' parameters $X_i$ in the BBN code:
\begin{itemize}
\item Gravitational constant $G_N$
\item Neutron lifetime $\tau_n$
\item Fine structure constant $\alpha$
\item Electron mass $m_e$
\item Average nucleon mass $m_N\equiv(m_n+m_p)/2$
\item Neutron-proton mass difference $Q_N\equiv m_n-m_p$
\item Binding energies of D, T, ${^3}$He, ${^4}$He, \lisi, \lise, ${^7}$Be
\end{itemize}
Clearly to vary each binding energy independently is unphysical, however our purpose is to determine the leading (linear) variation of all abundances, with respect to the code's input parameters, once and for all. Then given any specific theoretical model, we can construct a linear combination of variations of $X_i$
to account for any variation of a fundamental parameter. 

We should also consider the possibility that some states could change between bound to unbound, or stable and unstable, 
under a variation of some fundamental parameters. As discussed in Section~\ref{bindingvar}, we do not consider varying the binding energies $B_i$ of $A\leq 7$ nuclei past the point where the $Q$-value of any reaction relevant to standard BBN changes sign. If some $Q$-values approach zero we already find large variations in abundances. However, we should also check whether some reactions that are unimportant in standard BBN become relevant under a small change in parameters. If this were so then the linear approximation would quickly become inaccurate.

One could expect substantial changes in the final abundances of BBN if either the dineutron were bound, or the \beei\ nucleus were stable, at the time of nucleosynthesis. Such effects would not be seen in a purely linear analysis expanding about the present-day values.
We discuss the dineutron and \beei\ further in Section~\ref{n2beei}. We find that the dineutron binding cannot produce significant effects, given the other observational bounds on the variation of couplings. The requirement that \beei\ should remain unstable and short-lived during BBN gives a one-sided bound on the variation of $\alpha$, but only if light quark masses
are held constant relative to $\Lambda_c$. 

Our results for the nuclear response matrix are shown in Table \ref{dlnYdlnX}. The first thirteen rows constitute the transposed nuclear response matrix $C^T$. We also quote the dependence of the abundances on $\eta$ in the last row. Values are quoted to 2 d.\,p.\ or to 2 sig.\ fig.\ when the magnitude exceeds 1. Below we give a few comments concerning specific parameters $X_i$.

\begin{table}
\centering
\begin{tabular}{|c|c|c|c|c|c|}
\hline
$\partial\ln Y_a/\partial\ln X_i$  & D & \heth & \hefo & \lisi & \lise \\
\hline \hline
$G_N$         &  0.94 &  0.33 &  0.36 &  1.4  & -0.72 \\ \hline
$\alpha$      &  2.3  &  0.79 &  0.00 &  4.6  & -8.1  \\ \hline
$\tau_n$      &  0.41 &  0.15 &  0.73 &  1.4  &  0.43 \\ \hline
$m_e$         & -0.16 & -0.02 & -0.71 & -1.1  & -0.82 \\ \hline
$Q_N$         &  0.83 &  0.31 &  1.55 &  2.9  &  1.00 \\ \hline
$m_N$         &  3.5  &  0.11 & -0.07 &  2.0  &-12    \\ \hline
$B_{\rm D}$   & -2.8  & -2.1  &  0.68 & -6.8  &  8.8  \\ \hline
$B_{\rm T}$   & -0.22 & -1.4  &  0    & -0.20 & -2.5  \\ \hline
$B_{\rm 3He}$ & -2.1  &  3.0  &  0    & -3.1  & -9.5  \\ \hline
$B_{\rm 4He}$ & -0.01 & -0.57 &  0    &-59    &-57    \\ \hline
$B_{\rm 6Li}$ &  0    &  0    &  0    & 69    &  0    \\ \hline
$B_{\rm 7Li}$ &  0    &  0    &  0    &  0    & -6.9  \\ \hline
$B_{\rm 7Be}$ &  0    &  0    &  0    &  0    & 81    \\ \hline
\hline
$\eta$        & -1.6  & -0.57 &  0.04 & -1.5  &  2.1  \\ \hline
\end{tabular}
\caption{Response matrix $C$, dependence of abundances on nuclear parameters.} \label{dlnYdlnX}
\end{table}

\paragraph{Gravitational constant:}
A variation in $G_N$ or equivalently the Planck mass $M_P$, relative to $\Lambda_c$, affects the Hubble expansion. It is thus equivalent to changing the energy density of the Universe by a constant factor. During BBN this is also equivalent to adding or subtracting a number of relativistic species with radiation-like equation of state, the so-called ``effective number of neutrinos''; and also to the presence of a 'tracker' form of early dark energy \cite{Wetterich88}.

\paragraph{Neutron lifetime / \boldmath{$G_F$}:}
The neutron lifetime appears in normalising the $n\leftrightarrow p$ reaction rates and the free neutron decay. This is essentially the only important weak interaction, hence a variation in $\tau_n$ is equivalent to a variation of $G_F$.\footnote
{The variation of $\tau_n$ is also important in accounting for uncertainties in the laboratory value.} More precisely, for fixed $m_N$, $m_e$ and $Q_N$ and $g_a/g_v$ (the ratio of nucleon axial vector / vector couplings), there is a one-to-one relation between $\tau_n$ and $G_{\rm F}$ such that we can choose to work with one or the other parameter equivalently, using the mapping $\Delta \ln \tau_n = -2 \Delta \ln {G_F}$. 
Our results for variation of $\tau_n$ are then consistent with \cite{CocNunes06,MSW,ScherrerGN}.

\paragraph{Fine structure constant:}
Charged particle reactions incorporate an explicit $\alpha$-dependence in the Gamow factor, reflecting the leading effects of tunneling through the Coulomb barrier. Coulomb corrections also affect the $n\leftrightarrow p$ weak reaction rates. As noted in \cite{NollettLopez}, radiative reactions including $npd\gamma$ clearly have an additional dependence on $\alpha$. Reactions with charged particles in the final state also experience Coulomb suppression, though generally much less significant since the outgoing momenta are larger. For $Q>E$ the final state Coulomb suppression is negligible; the only cases where it may be worth considering are \heth(n,p)${^3}$H and \bese(n,p)\lise, with $Q$-values of $0.76\,$MeV and $1.64\,$MeV respectively. Such effects were estimated by Nollett and Lopez using the Coulomb wavefunctions for orbital angular momenta $l=0,1,2$ and appear to be small, the strongest dependence being $\vev{\sigma v} \propto \alpha^{-0.3}$. 

Our calculation of the $\alpha$-dependence as a {\em nuclear}\/ parameter does not include the effect of varying $\alpha$ on nucleon masses and nuclear binding energies; these are varied independently. The $\alpha$-dependence of binding energies will be incorporated at the second step through the response matrix $F$. On this second level we treat $\alpha$ as a fundamental coupling and incorporate the $\alpha$-dependence of the nuclear physics parameter $X_i$.

\paragraph{Neutron, proton and electron masses:}
We explicitly calculate the dependence of the $n\leftrightarrow p$ weak reactions on the electron mass $m_e$, and divide up the nucleon masses into an isospin-invariant part $m_N\equiv (m_p+m_n)/2$ and an isospin-violating part $Q_N\equiv m_p-m_n$. 
The dependence of final abundances on these nuclear parameters appears somewhat counterintuitive, since at this level we vary them with the condition that $\tau_n$ should remain constant. Thus effectively the variations in $m_e$ and $Q_{\rm N}$ are accompanied by compensating variations in the Fermi constant. When this is taken into account, our results for variation of $Q_{\rm N}$ are consistent with those of \cite{CocNunes06,MSW,KnellerLambda}. (However, our treatment of $m_e$ as a nuclear parameter differs from both \cite{MSW} and \cite{CocNunes06} hence the results are not directly comparable.)

The variation with respect to $m_N\equiv(m_n+m_p)/2$ is mainly due to the calculation of thermal reaction rates $\vev{\sigma v}$ at given temperature, since the relation between energy and velocity is affected; thus abundances which depend strongly on reaction rates will feel this variation. Again as with binding energies, we cannot at present tell to what extent reaction matrix elements depend on $m_N$ (with the exception of $npd\gamma$). However, for any choice of variations in fundamental parameters, we will see that the variation of $m_N$ is very small compared to other nuclear parameters, since we have fixed $\Lambda_c$ to be constant.

\subsection{Variation of binding energies} \label{bindingvar}
Table~\ref{dlnYdlnX} indicates that by far the largest sensitivity of abundances to nuclear parameters involves the variation of \hefo, \lise\ and \bese\ binding energies. This can easily be understood since the reaction rate of \heth$(\alpha,\gamma)$\bese\ with $Q$-value 1.59\,MeV is very sensitive to changes in these (numerically large) binding energies. 
In order to have a possibly more intuitive idea of the sensitivity of $Y_a$ to binding energies 
we display in Table~\ref{tabYb} the dependence of abundances on the binding energy per nucleon $b_i\equiv B_i/A_i$, that is, $\partial \ln Y/\partial b_i$, since it is more meaningful to compare values of $b_i$ between nuclei of different $A$. 
\begin{table} 
\centering
\begin{tabular}{|c|c|c|c|c|c|}
\hline
$\partial \ln Y_a/\partial X_i$ & D & \heth & \hefo & \lisi & \lise \\
\hline\hline
$b_{\rm D}  $ & -2.5  & -1.9  & 0.61 & -6.1  &  7.9  \\ \hline
$b_{\rm T}  $ & -0.08 & -0.50 & 0    & -0.07 & -0.88 \\ \hline
$b_{\rm 3He}$ & -0.83 &  1.2  & 0    & -1.2  & -3.7  \\ \hline
$b_{\rm 4He}$ &  0    & -0.08 & 0    & -8.3  & -8.0  \\ \hline
$b_{\rm 6Li}$ &  0    &  0    & 0    & 13    &  0    \\ \hline
$b_{\rm 7Li}$ &  0    &  0    & 0    &  0    & -1.2  \\ \hline
$b_{\rm 7Be}$ &  0    &  0    & 0    &  0    & 15    \\ \hline
\end{tabular}
\caption{Dependence of abundances on binding energy per nucleon, in units of MeV$^{-1}$.}\label{tabYb}
\end{table}
We see that the dependence of the lithium abundances on the \hefo, \lisi\ and \bese\ binding energies is still dominant.

Each reaction $Q$-value is determined by the masses of reactants and products: the relevant parameters are the binding energies of nuclei up to $A=7$. The $Q$-value of each reaction affects the abundances via the reverse thermal reaction rate relative to the forward rate, and via phase space and radiative emission factors in the reaction cross-sections. 

The reverse reaction rate is simply related to the forward rate via statistical factors, due to time reversal invariance (see for example \cite{NACRE}): the relevant dependence is  
\beq
\frac{\vev{\sigma v}_{34\rightarrow 12}}{\vev{\sigma v}_{12\rightarrow 34}} \propto e^{-Q/T}.
\eeq
The $Q$-dependence of radiative capture reactions (assuming a dominant electric dipole) is 
\beq \label{radiative}
\sigma(E) \propto E_\gamma^3 \sim (Q+E)^3
\eeq
whereas for $2\rightarrow 2$ inelastic scattering or transfer reactions the dependence is
\beq \label{2to2}
\sigma(E) \propto \beta \sim (Q+E)^{1/2}
\eeq
where $\beta$ is the outgoing channel velocity. In the current treatment we assume $E\ll Q$ at relevant temperatures, and simply scale rates by the appropriate power of $Q$. Clearly this breaks down when $Q$ approaches zero, and we have not considered varying any binding energy to the point where this happens. A more accurate treatment would involve applying the phase space dependences directly to the cross-sections, for example in the S-factor description of charged particle reactions, which involves an expansion in $E$; the dependence on $(Q+E)$ can then be applied order by order.

For resonances, which appear in reaction rates as terms varying with $e^{-E_r/T}$, we scale their contributions by the appropriate power of $(Q+E_r)$.\footnote{The resulting variations of abundances are indistinguishable from the result of scaling by a power of $Q$, within our uncertainties.} The question of whether $E_r$ should be varied is still open. The minimal assumption is that the mass-energy of the resonance varies with the mass-energy in the incoming channel, thus $E_r$ would be constant to first approximation.

A variation in binding energies can have two kinds of effect. The $Q$ value can change the time when a reaction drops out of equilibrium, for instance the $n+p \rightarrow d+\gamma$ reaction. Or it can change the absolute rate of a reaction, and thus the production rate of a given species, for example the \bese-producing reaction whose cross-section varies with $Q^3$.

Whether the reaction matrix elements have a dependence on the binding energies and on $Q$ is in general not clear because there is no systematic effective theory for multi-nucleon reactions. The exception is the $npd\gamma$ reaction, for which we have implemented a nuclear effective theory result \cite{Chen99} where dependence on $B_{\rm D}$ is explicit. Other parameters of this effective theory may also have some quark mass dependence, and there are ongoing efforts to connect it to chiral perturbation theory and thus to QCD parameters, specifically $m_q/\Lambda_c$ \cite{Chenprivate}. 

It has been suggested (see \cite{KnellerLambda}) that the cross-sections of other reactions involving deuterium, specifically when D appears in the ingoing channel, will have a significant additional dependence on $B_{\rm D}$. To first order, the scattering length varies as $B_{\rm D}^{-1/2}$, and cross-sections in the low-energy limit may be expected to vary with $B_{\rm D}^{-1}$, in addition to previously discussed effects. In Table \ref{dlnYdlnX} we present values for the dependence on $B_D$, assuming this additional dependence {\em not}\/ to be present. When, however, a $\sigma\propto 1/B_{\rm D}$ dependence is implemented for the $d(p,\gamma)$\heth, $d(d,n)$\heth\ and $d(d,p)t$ reactions, we obtain
\beq
 \frac{\partial\ln Y_a}{\partial\ln B_{\rm D}} (\rm D, \mbox{\heth}, \mbox{\hefo}, \mbox{\lisi}, \mbox{\lise}) = (-1.5, -2.4, 0.66, -5.6, 7.5).
\eeq
The resulting uncertainties in reaction rates have a subleading effect on abundances, compared to other effects of varying $B_{\rm D}$. To be consistent we should similarly consider the effects of all other binding energies on reaction cross-sections beyond the kinematic factor of Eqs.~(\ref{radiative},\ref{2to2}); however in any case it is not clear whether the scattering length is the correct parameter to consider. Therefore we will take the case without this $\sigma\propto 1/B_{\rm D}$ dependence as our final result; the other case serves as an illustration of possible further effects of binding energies on reaction rates.

Our results for the dependence on $B_{\rm D}$ are consistent with those of \cite{Chamoun05,MSW,KnellerLambda} (allowing for the different treatments of D-destroying reactions) but differ from \cite{CocNunes06} and \cite{Dmitriev03} for the dependence of the D and \lise\ abundances. The discrepancy with respect to those works arises from a different treatment of the $npd\gamma$ reaction. It is also not clear to us if the effect of changing $B_{\rm D}$ on other reaction $Q$-values and rates was included.

\subsection{Nuclear rates}
In order to estimate which reactions are more or less important in the variation of the final abundances, we varied each thermal averaged cross-section $\vev{\sigma v}$ by a temperature-independent factor, preserving the relation between forward and reverse rates. 
The aim is to diagnose which reactions one should focus on in discussing the sensitivity of observed abundances to variations of couplings.
The $n\leftrightarrow p$ weak interactions influence every abundance nontrivially (we will treat them analytically); in addition, based on our results, we designated certain other reactions as important. These are given in Table \ref{importants}.\footnote{As usual in BBN simulations, the slow $\beta$-decays of tritium and \bese\ are accounted for by adding on the T and \bese\ abundances to \heth\ and \lise\ respectively at the end of the run, when other nuclear reactions have frozen out.}
\begin{table}
\centering
\begin{tabular}{|l|c|c|c|c|c|c|}
\hline
Reaction  &    $Q$ value [MeV] & D & \heth & \hefo & \lisi & \lise \\
\hline\hline
$p(n,\gamma)d$	 	& 2.22 & -0.2 &  0.1 &  0 & -0.2 &  1.3 \\ \hline
$d(p,\gamma)$\heth 	& 5.49 & -0.3 &  0.4 &  0 & -0.3 &  0.6 \\ \hline
$d(d,n)$\heth 		& 3.27 & -0.5 &  0.2 &  0 & -0.5 &  0.7 \\ \hline
$d(d,p)t$ 		& 4.03 & -0.5 & -0.3 &  0 & -0.5 &  0.1 \\ \hline
$d(\alpha,\gamma)$\lisi & 1.47 &  0   &  0   &  0 &  1.0 &  0   \\ \hline
\heth$(n,p)t$		& 0.76 &  0   & -0.2 &  0 &  0   & -0.3 \\ \hline
\heth$(d,p)$\hefo	& 18.35&  0   & -0.8 &  0 &  0   & -0.7 \\ \hline
\heth$(\alpha,\gamma)$\bese & 1.59& 0 &  0   &  0 &  0   &  1.0 \\ \hline   
\lisi$(p,\alpha$)\heth 	& 4.02 &  0   &  0   &  0 & -1.0 &  0   \\ \hline
\bese$(n,p)$\lise  	& 1.64 &  0   &  0   &  0 &  0   & -0.7 \\ \hline
\end{tabular}
\caption{Leading dependence of abundances on thermal averaged cross-sections $\partial \ln Y_a/\partial \ln \vev{\sigma v}_i$ for important reactions (1 d.\,p.)} \label{importants}
\end{table}
We judge a reaction cross-section to be important if the dependence of any abundance on any given reaction cross-section $\partial\ln Y_a/\partial\ln \vev{\sigma v}_i$ is more than $0.1$. In every case the dependences are order unity or smaller, and for many ``important'' reactions only a few abundances are significantly affected. A reaction cross-section can be unimportant (for our purposes) for one of two reasons: either it is so small that the reaction is irrelevant, or the reaction is so rapid, compared to the Hubble rate and to other slower reactions, that the rates of change of abundances are almost independent of the cross-section.
Hence our list of ``important'' reactions differs from 
\cite{SKM93}.

Note also that the \hefo\ abundance does not depend on any nuclear reaction cross-section (apart from $n\leftrightarrow p$). Furthermore, the reactions involving \lisi\ are not important for any other measurable primordial abundance.

In implementing the variations of nuclear parameters as displayed in Table~\ref{dlnYdlnX}, we do not directly use the effects of a temperature-independent variation of integrated cross-section $\vev{\sigma v}$ given in Table~\ref{importants}. Since variations of $X_i$ in general result in temperature-dependent variations of reaction rates, we implement them directly within the integration code.

\section{Relations to fundamental parameters} \label{fundamentals}
The next task is to connect these nuclear parameters to fundamental parameters $G_k$ at a higher energy scale. We consider the following six fundamental parameters $G_k$:
\begin{itemize}
\item Gravitational constant $G_N$
\item Fine structure constant $\alpha$
\item Electron mass $m_e$
\item Light quark mass difference $\delta_q \equiv m_d-m_u$
\item Averaged light quark mass $\hat{m}\equiv (m_d+m_u)/2 \propto m_\pi^2$
\item Higgs v.e.v.\ $\vev{\phi}$.
\end{itemize}
An additional parameter is the strange quark mass $m_s$. We have omitted it from our list because the present theoretical uncertainties of how it influences the nuclear parameters are too high. 

The dependence of the fundamental parameters $G_k$ on the nuclear parameters $X_i$ is encoded in the matrix $F$ defined in Eq.~(\ref{Fdef}): our estimates of $F$ are shown in Table \ref{dlnXdlnG}.
\begin{table}
\centering
\begin{tabular}{|c||c|c|c|c|c|c|}
\hline 
$\partial \ln X_i/\partial\ln G_k$ & $G_{\rm N}$ & $\alpha$ & $\vev{\phi}$ & $m_e$ & $\delta_q$ & $\hat{m}$ \\
\hline \hline
$G_{\rm N}$ & 1 & 0 & 0 & 0 & 0 & 0 \\
\hline 
$\alpha$ & 0 & 1 & 0 & 0 & 0 & 0 \\
\hline
$\tau_n$ & 0 & 3.86 & 4 & 1.52 & -10.4 & 0 \\
\hline
$m_e$ & 0 & 0 & 0 & 1 & 0 & 0 \\
\hline
$Q_N$ & 0 & -0.59 & 0 & 0 & 1.59 & 0 \\
\hline
$m_N$ & 0 & 0 & 0 & 0 & 0 & 0.048 \\
\hline
$B_{\rm D}$ & 0 & -0.0081 & 0 & 0 & 0 & $-4$ \\
\hline
$B_{\rm T}$ & 0 & -0.0047 & 0 & 0 & 0 & $-2.1 f_{\rm T}$ \\
\hline
$B_{3\rm He}$ & 0 & -0.093 & 0 & 0 & 0 & $-2.3 f_{3\rm He}$ \\
\hline
$B_{4\rm He}$ & 0 & -0.030 & 0 & 0 & 0 & $-0.94 f_{4\rm He}$ \\
\hline
$B_{6\rm Li}$ & 0 & -0.054 & 0 & 0 & 0 & $-1.4 f_{6\rm Li}$ \\
\hline
$B_{7\rm Li}$ & 0 & -0.046 & 0 & 0 & 0 & $-1.4 f_{7\rm Li}$ \\
\hline
$B_{7\rm Be}$ & 0 & -0.088 & 0 & 0 & 0 & $-1.4 f_{7\rm Be}$ \\
\hline
\end{tabular}
\caption{Response matrix $F$, dependence of nuclear parameters $X_i$ on fundamental parameters $G_k$}\label{dlnXdlnG}
\end{table}
We now discuss how to derive the entries of this matrix.

The gravitational constant enters as before, corresponding to the $(1,1)$ entry of unity in Table~\ref{dlnXdlnG}. The fine structure constant influences the abundances in two ways. First we have the direct influence of its variation while keeping other nuclear parameters fixed. This is accounted for by the value $1$ in the $(2,2)$ element. Secondly, it influences the nucleon masses and nuclear binding energies. This yields the other elements in the second column of Table~\ref{dlnXdlnG}. As a nuclear parameter $\alpha$ does not depend on $\vev{\phi}$, $m_e$, {\it etc.}, as reflected in the second row of Table~\ref{dlnXdlnG}.

The $\alpha$-dependence of the neutron lifetime $\partial \ln \tau_n = 3.86\, \partial \ln \alpha$ enters via the $n$-$p$ mass difference. The full dependence of $\tau_n$ on fundamental parameters can be found in \cite{MSW}. Similarly as for $\alpha$, the electron mass as a fundamental parameter has an effect on the neutron lifetime, $\partial \ln \tau_n = 1.52\, \partial \ln m_e$, and thus also on $n\leftrightarrow p$ reaction rates.

Elementary $u$ and $d$ quark masses appear as fundamental parameters. They may be divided up as $(m_u+m_d)/2$, which influences the pion mass, which in turn determines the behaviour of nuclear forces; and $m_d-m_u$, which influences the neutron-proton mass difference. The average light quark mass also affects the nucleon mass $m_N$ via the so-called sigma term; this holds also for the strange quark mass through the ``strangeness content'' of the nucleon. The effect of varying pion ({\it i.e.}\ quark) mass on nuclear binding energies and reaction cross-sections is in general not known. However, for the binding energy of deuterium and the $npd\gamma$ reaction the dependence has been found to some approximation using effective theories.

At this stage we vary the Higgs v.e.v.\ {\em independently}\/ of the elementary fermion masses, thus it only influences weak reactions via $G_F$: the relevant reactions are neutron decay and $n\leftrightarrow p$.

\subsection{Fine structure constant}
The dependence on $\alpha$ of the masses of (composite) particles and nuclear binding energies is found by estimates of electromagnetic self-energy or binding energy, for example \cite{GasserLeutwyler} for the proton-neutron mass difference, and the semi-empirical mass formula for nuclei. Note that the fractional variation of $m_N$ with $\alpha$ is negligibly small. More precise estimates for nuclear binding energies have been made using realistic models of nuclear forces (see \cite{Pudliner97} and \cite{Pieper01}) and similar values 
appear in \cite{NollettLopez}.
Variation in $\alpha$ as a fundamental parameter leads to variations of nuclear parameters:
\begin{multline}
 \Delta \ln (
 \tau_n, Q_N, B_{\rm D}, B_{\rm T}, B_{{3}\rm He}, B_{{4}\rm He}, B_{{6}\rm Li}, B_{{7}\rm Li}, B_{{7}\rm Be}) = \\
 (3.86, -0.59, -0.0081, -0.0047, -0.093, -0.030, -0.054, -0.046, -0.088) \Delta \ln \alpha.
\end{multline}
Then the resulting variations of abundances are 
\beq
 \Delta \ln (Y_{\rm D}, Y_{3\rm He}, Y_{4\rm He}, Y_{6\rm Li}, Y_{7\rm Li}) = 
 (3.6,0.95,1.9,6.6,-11) \Delta \ln \alpha
\eeq
These results are similar to those of \cite{BergstromIR} and \cite{NollettLopez}. The \lise\ abundance here is more sensitive to $\alpha$ than the Nollett \& Lopez estimate \cite{NollettLopez}. This appears due to their use of a cluster model for the \heth$(\alpha,\gamma)$\bese\ reaction, whereas
our treatment of this reaction simply uses the Gamow factor. 

\subsection{Electron mass}
The electron mass affects the neutron lifetime 
via
\[ 
\frac{\partial \ln \tau_n}{\partial \ln m_e} \simeq 1.52.
\]
Adding this variation to the effects already calculated gives, for the variation of $m_e$ as fundamental parameter:
\beq
\Delta \ln (Y_{\rm D}, Y_{3\rm He}, Y_{4\rm He}, Y_{6\rm Li}, Y_{7\rm Li}) = (0.46,0.21,0.40,0.97,-0.17) \Delta \ln m_e.
\eeq
Our result for the variation of $m_e$ as fundamental parameter is very similar to the semi-analytic result of \cite{MSW}.

\subsection{Pion and light quark masses} \label{qcdstuff}
The effect of light quark masses on the nucleon masses can be found by considering low-energy hadron physics (see for example \cite{GasserLeutwyler, BorasoyMeissner}). We consider an average light quark mass $\hat{m}\equiv (m_u+m_d)/2$, the mass difference $\delta_q \equiv m_d-m_u$ and the strange mass $m_s$. Note that the $m_s$ dependence of hadronic physics is still quite unclear, for example the strangeness content of the nucleon is subject to at least 50\% uncertainty. The nucleon mass gets a contribution from nonzero quark masses
\beq
 m_N = m_N^{(0)}(\Lambda_c) + \sigma_{\pi N} (\hat{m})
\eeq 
where $m_N^{(0)}$ is the mass in the chiral limit and 
\beq \label{dmNdmhat}
\sigma_{\pi N} = \frac{\hat{m}}{2m_N}\vev{p|\bar{u}u+\bar{d}d|p}.
\eeq
For $\vev{p|\bar{u}u+\bar{d}d|p}$ depending only on $\Lambda_c$, one has $\partial \ln \sigma_{\pi N}/\partial \ln \hat{m} = 1$ and therefore
\beq
\frac{\partial \ln m_N}{\partial \ln \hat{m}} = \frac{\sigma_{\pi N}}{m_N}  \simeq 0.048
\eeq
and for the strange quark
\beq \label{dmNdms}
y\equiv \frac{2\vev{p|\bar{s}s|p}}{\vev{p|\bar{u}u+\bar{d}d|p}}\qquad \Rightarrow \qquad
\frac{\partial \ln m_N}{\partial \ln m_s} = \frac{m_s}{\hat{m}}\frac{y\sigma_{\pi N}}{2m_n}
\simeq 0.12 \pm 0.12
\eeq
given $m_s/\hat{m}\simeq 25$ and $y=0.2\pm 0.2$ \cite{strangelat05}. 

The strange quark mass is close to the nonperturbative scale $\Lambda_c$. For this reason a systematic or physically meaningful treatment of the dependence of nuclear quantities on $m_s$ has not been possible so far. 
Thus our results have the caveat that variation in $m_s/\Lambda_c$ is not yet accounted for.

The dependence of the nucleon mass difference $Q$ on quark masses was estimated in \cite{GasserLeutwyler}:
in units where $\Lambda_c$ is constant, we have
\beq
\Delta Q_N \simeq (-0.76 \Delta \ln \alpha +2.05 \Delta \ln \delta_q)\,{\rm MeV}
\qquad \Rightarrow \qquad \frac{\Delta \ln Q}{\Delta \ln \delta_q} 
\simeq 1.59.
\eeq
Recent lattice QCD studies with dynamical quarks \cite{Beane06lattice} have calculated the dependence of the nucleon mass splitting on $\delta_q$: the result is consistent with the estimate we adopt.

The pion mass is crucial for nuclear forces and its leading dependence on $\hat{m}$ follows from chiral perturbation theory as
\beq
 m_{\pi}^2 = \hat{m}\vev{\bar{u}u+\bar{d}d} f_\pi^{-2}.
\eeq
Here we assume the leading order where $f_\pi$ and $\vev{\bar{u}u+\bar{d}d}$ depend only on $\Lambda_c$, therefore 
$\Delta \ln m_\pi \simeq \frac{1}{2} \Delta \ln \hat{m}$. 

Static properties of nuclei, and importantly for BBN, nuclear binding energies, depend strongly on the pion mass, which determines the range of attractive nuclear forces. Quantum Monte Carlo calculations have been performed with realistic nuclear potentials and accurately reproduce many experimental properties \cite{Pieper01, Pudliner97}. One-pion exchange and two-pion exchange are dominant contributions within the expectation values of the two- and three-nucleon potentials respectively. Currently such studies have not been extended to determine the functional dependence of binding energies on the pion mass in general. This dependence would in any case have uncertainties due to subleading effects of pion mass (or equivalently light quark masses) on other terms in the nucleon-nucleon potential \cite{BeaneSavage02}.

However, the dependence of the deuteron binding energy on the pion mass has been extensively studied within low-energy effective theory \cite{EpelbaumMeissner02,BeaneSavage02}: the result may be expressed as 
\beq
 \Delta \ln B_{\rm D} = r \Delta \ln m_\pi = \frac{r}{2} \Delta \ln \hat{m}
\eeq
for small variations about the current value \cite{YooScherrer}, with $-10\leq r \leq -6$.\footnote{Our definition of $r$ differs by a sign from \cite{YooScherrer}.} We will also take this dependence as a guide for the likely pion mass dependence of other binding energies. Although the size of the deuteron binding appears due to an accidental cancellation between attractive and repulsive forces, its derivative with respect to $m_\pi$ (which is just $B_{\rm D}/m_\pi$ times $r$) is not expected to be subject to any cancellation.
We also expect that the pion contribution to the total binding energy should increase with the number of nucleons; a proportionality to $(A-1)$ seems reasonable to obtain correct scaling at both small and large $A$. Hence to estimate the effect of pion mass on the binding energy of a nucleus $B_i$ we set
\beq \label{dBdmpi}
 \frac{\partial B_i}{\partial m_\pi} = f_i (A_i-1) \frac{B_{\rm D}}{m_\pi} r \simeq
 -0.13 f_i (A_i-1) 
\eeq
taking $r\simeq-8$. The numerical constants $f_i$ are expected to be of order unity, but will differ between light nuclei due to peculiarities of the shell structure, {\em etc}. Our normalization corresponds to $f_{\rm D}=1$. 
Then the nontrivial dependences of nuclear parameters on $\hat{m}$ are
\begin{multline}
 \Delta \ln (B_{\rm D}, B_{\rm T}, B_{3\rm He}, B_{4\rm He}, B_{6\rm Li}, B_{7\rm Li}, B_{7\rm Be}, m_N) 
 \simeq \\
 (0.5r, 0.26f_{\rm T}r, 0.29f_{3\rm He}r, 0.12f_{4\rm He}r, 0.17f_{6\rm Li}r, 
 0.17f_{7\rm Li}r, 0.18f_{7\rm Be}r, 0.048) \Delta \ln \hat{m}.
\end{multline}
For the $\hat{m}$ dependence of abundances due to the variation of binding energies we then have
\beq
 \left.\frac{\partial \ln Y_a}{\partial \ln \hat{m}}\right|_{B} = \frac{r}{2}
 \sum_i f_i \frac{(A_i-1)B_{\rm D}}{B_i}
 \frac{\partial \ln Y_a}{\partial \ln B_i}. 
\eeq
Taking account also of the small effect of $\hat{m}$ on the nucleon mass $m_N$, the resulting dependence of abundances on $\hat{m}$ is 
\bea
 \frac{\partial \ln Y_{\rm D}}{\partial \ln \hat{m}} &\simeq& 
 11 
 + 0.5 f_T + 5 f_{3\rm He} \nonumber \\
 \frac{\partial \ln Y_{3\rm He}}{\partial \ln \hat{m}} &\simeq& 
 8 
 + 3 f_T - 7 f_{3\rm He} \nonumber \\
 \frac{\partial \ln Y_{4\rm He}}{\partial \ln \hat{m}} &\simeq& 
 -2.7 
 \nonumber \\
 \frac{\partial \ln Y_{6\rm Li}}{\partial \ln \hat{m}} &\simeq& 
 27 
 + 0.4 f_{\rm T} + 7 f_{\rm 3He} + 55 f_{4\rm He} - 96 f_{6\rm Li} \nonumber \\
 \frac{\partial \ln Y_{7\rm Li}}{\partial \ln \hat{m}} &\simeq& 
 - 36 
 + 5 f_{\rm T} + 22 f_{3\rm He} + 54 f_{4\rm He} + 9 f_{7\rm Li} -115 f_{7\rm Be}.
\eea
Even if we consider that some contributions could cancel against one another due to the values of the $f_i$, the magnitude of these variations is striking, particularly concerning the lithium abundances. To get an idea of the possible effect of cancellations, we may set all $f_i$ to unity and find the dependences
\beq
\Delta \ln (Y_{\rm D}, Y_{3\rm He}, Y_{4\rm He}, Y_{6\rm Li}, Y_{7\rm Li}) \simeq 
(17, 5, -2.7, -6, -61)
\Delta \ln \hat{m}.
\eeq
One may also consider to what extent varying $\hat{m}$ or the pion mass may affect reaction cross-sections beyond the $npd\gamma$ reaction. It seems very likely that matrix elements would acquire nontrivial dependence on $m_\pi$; however, since the dependence of abundances on reaction cross-sections is relatively mild (see Table~\ref{importants}),
the dependence via reaction matrix elements is unlikely to compete with the very large effects arising through the variation of binding energies.

\subsection{Stability of dineutron and \beei} \label{n2beei}
The effects of a bound dineutron on BBN were studied in \cite{Knellern2}, where it was found that the final abundances were essentially unaffected as long as the dineutron binding energy remained smaller than $B_{\rm D}$. An effective field theory analysis may also be applied to the binding of the dineutron, which lies in a different channel from the deuteron \cite{BeaneSavage02}. The sensitivity of the dineutron system to the pion mass, or equivalently to $\hat{m}$, is found to be comparable to that of the deuteron. It is very unlikely that the binding of the dineutron could become significant, since the fractional variations of $B_{\rm D}$ for the range of variations of couplings considered in this paper are at most of order 5\%, which amounts to $\Delta B_{\rm D}\simeq 0.1\,$MeV (see Section~\ref{sepbounds}). In earlier work a simple potential model was used \cite{DentF} to investigate the deuteron and dineutron binding, with a similar result: the variation in the quark mass parameter $\hat{m}/\Lambda_c$ required to bind the dineutron is much larger than that required to cause even a 100\% variation in $B_{\rm D}$. 

The decay of \beei\ to two \hefo\ nuclei has a $Q$-value of only $0.092\,$MeV, thus it is conceivable that even a very small variation of parameters could result in a stable (or long-lived) \beei\ at the time of BBN. This would have dramatic consequences for the abundances of heavier nuclei since it would then be possible to synthesize carbon directly by two-body reactions. Hence we can immediately rule out any variation that causes this $Q$-value to change sign.

Electromagnetic contributions to the \beei\ binding are estimated from the results of quantum Monte Carlo calculations \cite{Pieper01} as before. We find
{$
 \Delta \ln B_{\rm 8Be} \simeq -0.0588 \Delta \ln \alpha,
$} 
thus the $Q$-value varies as 
\beq
 \Delta Q_{8\rightarrow2\alpha} \simeq (1.60\,\mbox{MeV}) \Delta \ln \alpha.
\eeq
Large negative values of $\Delta \alpha$ are excluded: we find a firm limit $\Delta \ln \alpha \geq -5.7\%$, under the condition that other fundamental parameters that may affect the sign of $Q$, {\it i.e.}\ quark masses, do not vary with respect to $\Lambda_c$. However, we will obtain stronger limits on the variation of $\alpha$ by considering the response of the observed BBN abundances: see Section~\ref{sepbounds}.

The $Q$-value for \beei\ decay results from a nearly exact cancellation between \beei\ and \hefo\ binding energies, thus its dependence on light quark masses is subject to very large theoretical uncertainty. We again estimate the $\hat{m}$-dependence 
in terms of the deuteron binding and find $\Delta \ln B_{\rm 8Be} = -1.1 f_{\rm 8Be}\, \Delta \ln \hat{m}$. Thus for the $Q$-value,
\beq
\Delta Q \simeq (\left[9 + 62 (f_{\rm 8Be} - 1) - 53 (f_{\rm 4He} - 1) \right]\mbox{MeV}) \Delta \ln \hat{m}.
\eeq
The expression inside brackets depends very strongly on the unknown $f_i$ factors: its likely magnitude is $10$--$50\,$MeV barring cancellations. We would then have a one-sided bound on variation of $\hat{m}$ at the 1\% level or better from stability of \beei. At present we do not know the sign of the prefactor and the use of \beei\ to place useful bounds on the variation of QCD parameters must be left to future work.

\section{Dependence of abundances on fundamental parameters}

\subsection{Fundamental response matrix}
We next combine the nuclear response matrix, Table~\ref{dlnYdlnX} with the relations between nuclear and fundamental parameters (Table~\ref{dlnXdlnG}), according to Eq.~(\ref{RequalsCF}).
Table~\ref{dlnYdlnG} shows the resulting dependences of abundances on fundamental parameters, as encoded in the matrix $R$. This table is our central result.
\begin{table}
\centering
\begin{tabular}{|c||c|c|c|c|c|}
\hline 
$\partial \ln Y_a/\partial\ln G_k$ &  D   & \heth & \hefo & \lisi  & \lise \\ \hline \hline
$G_{\rm N}$                        & 0.94 & 0.33  & 0.36  &  1.4   & -0.72 \\ \hline 
$\alpha$                           & 3.6  & 0.95  & 1.9   &  6.6   &-11    \\ \hline
$\vev{\phi}$                       & 1.6  & 0.60  & 2.9   &  5.5   &  1.7  \\ \hline
$m_e$                              & 0.46 & 0.21  & 0.40  &  0.97  & -0.17 \\ \hline
$\delta_q$                         &-2.9  &-1.1   &-5.1   & -9.7   & -2.9  \\ \hline
$\hat{m}$                          &17    & 5.0   &-2.7   & -6     &-61    \\ \hline
\hline
$\eta$                             &-1.6  &-0.57  & 0.04  & -1.5   &  2.1  \\ \hline
\end{tabular}
\caption{Response matrix $R$, dependence of abundances $Y_i$ on fundamental parameters $G_k$}\label{dlnYdlnG}
\end{table}

In treating the $\hat{m}$-dependences, which arise from the nuclear binding energies with their uncertain values of $f_i$, we have given the values which arise when setting all $f_i$ to unity. Alternatively, if all the $f_i$ are of order unity but the term with the largest prefactor dominates, we would obtain\footnote{If, on the contrary, there is substantial cancellation then the dependences of the deuterium, \heth\ and lithium abundances on $\hat{m}$ may be smaller.}
\beq
\frac{\partial \ln Y}{\partial\ln \hat{m}} (\rm D, \mbox{\heth}, \mbox{\hefo}, \mbox{\lisi}, \mbox{\lise}) = (11, 8.4, -2.7, -96 f_{6Li}, -115 f_{7Be}).
\eeq

The dependence on $G_{\rm N}$ is consistent with the results of \cite{ScherrerGN,MSW} and \cite{Chamoun05,Landau04}, once one translates from units where $G_{\rm N}$ is constant to ours where $\Lambda_c$ is constant.

The \hefo\ dependence was previously calculated in \cite{MSW} by semi-analytic methods: our results for the dependence on fundamental parameters are similar.

The $\vev{\phi}$ dependence may be compared to the result of \cite{Landau04} for the variation of $G_{\rm F}\propto \vev{\phi}^{-2}$ (using a semi-analytic method): the results for \hefo, \lisi\ and \lise\ are similar, but much smaller variations of D and \heth\ are obtained in \cite{Landau04}.

\subsection{Bounds on separate variations of fundamental couplings} \label{sepbounds}
The first application of the result is in setting bounds on the variation of each fundamental parameter considered separately, under the assumption that only one parameter varies at once. We may consider three observational determinations of primordial abundances (see Appendix~\ref{AppendixB}): deuterium, \hefo\ and \lise. However, the observed \lise\ abundance deviates by a factor two to three from the value predicted by standard BBN theory (SBBN), and systematic uncertainties related to stellar evolution exist \cite{Korn06}. Thus, we use the former two, D and \hefo, to constrain the allowed variations of the fundamental constants individually.
For deuterium we take $2 \sigma$ limits;
for \hefo\ we consider instead the ``conservative allowable range'' of \cite{OliveSkillman04}. The resulting constraints are given in Table~\ref{AllowedVariations}.
\begin{table}
\centering
\begin{tabular}{|clcc|}
\hline
$-19 \% $&$\le\ \Delta \ln G_N        $&$\le$&$ +10 \%  $\\ \hline
$-3.6\% $&$\le\ \Delta \ln \alpha     $&$\le$&$ +1.9\%  $\\ \hline
$-2.3\% $&$\le\ \Delta \ln \vev{\phi} $&$\le$&$ +1.2\%  $\\ \hline
$-17 \% $&$\le\ \Delta \ln m_e        $&$\le$&$ +9.0\%  $\\ \hline
$-0.7\% $&$\le\ \Delta \ln \delta_q   $&$\le$&$ +1.3\%  $\\ \hline
$-1.3\% $&$\le\ \Delta \ln \hat{m}    $&$\le$&$ +1.7\%  $\\
\hline
\end{tabular}
\caption{Allowed individual variations ($2\sigma$ or ``conservative allowable range'') of fundamental couplings}\label{AllowedVariations}
\end{table}

\subsection{Sensitivity matrix}
The relative precision of the observational determination of primordial abundances is best for \hefo\ and somewhat poorer for D and \lise. This can be taken into account by defining a ``sensitivity matrix'' $S$ with elements $s_{ak}$:
\beq
s_{ak} = \sigma_a^{-1} \frac{ \partial Y_a}{\partial \ln G_k} 
\simeq \frac{Y_{a,\,\rm th}}{\sigma_a} r_{ak,\,\rm th}
\eeq
where the subscript {\small `th'} denotes a quantity evaluated about the values obtained in SBBN, and the approximation follows from taking a linear dependence. Here $\sigma_a$ is the 1$\sigma$ error of the observational determination of the primordial abundance $Y_a$: thus $Y_a/\sigma_a$ is a measure of the precision with which a given abundance is known. The current situation of theory and observation is summarized in Appendix~\ref{AppendixB}. We adopt SBBN theoretical values
\bea
Y_{\rm D,th} &=& 2.61 \times 10^{-5} \nonumber \\ 
Y_{\rm 4He,th} &=& 0.2478 \nonumber \\
Y_{\rm 7Li,th} &=& 4.5 \times 10^{-10} 
\eea
and 1$\sigma$ observational errors
\bea
\sigma_{\rm D} &=& 0.4 \times 10^{-5} \nonumber \\ 
\sigma_{\rm 4He} &=& 0.009 \nonumber \\
\sigma_{\rm 7Li} &=& 0.5 \times 10^{-10}. 
\eea
We then obtain the sensitivity matrix $S$ given in table~\ref{SensitivityMatrix}.
\begin{table}
\centering
\begin{tabular}{|c||c|c|c|}
\hline
$\sigma_a^{-1}\partial Y_a/\partial\ln G_k$ &  D  & \hefo & \lise \\ \hline
\hline
$G_{\rm N}$                        & 6.1 & 9.9  & -6.5   \\ \hline
$\alpha$                           & 24  & 52   & -100   \\ \hline
$\vev{\phi}$                       & 10  & 79   &  15    \\ \hline
$m_e$                              & 3.0 & 11   & -1.5   \\ \hline
$\delta_q$                         & -19 & -140 & -26    \\ \hline
$\hat{m}$                          & 110 & -74  & -550   \\ \hline
\end{tabular}
\caption{Sensitivity matrix $S$}\label{SensitivityMatrix}
\end{table}

The matrix elements $s_{ak}$ have a simple interpretation. If we perform a variation of a given fundamental coupling $G_k$ by $1\%$ while keeping all other fundamental couplings fixed, the variation of $Y_a/\sigma_a$ is given by $s_{ak}/100$. The range of variation of each $G_k$ corresponding to the 1$\sigma$ ({\it etc.})\ range of any observed abundance can easily be read off. Thus for example the currently allowed 1$\sigma$ range of $Y_{\rm 4He}$ \cite{OliveSkillman04}
corresponds to a variation in $\alpha$ of $1/52 \simeq 1.9\%$ each side of a central value, or $3.8\%$ in total.
Considering individual variation of couplings we find that at present the \hefo\ abundance is the most sensitive measurement of the couplings $G_{\rm N}$, $\phi$, $m_e$ and $\delta_q$; on the other hand $\alpha$ and the average light quark mass $\hat{m}$ are most limited by \lise.

\subsection{Inverse sensitivity matrix}

A quantity like the inverse of the sensitivity matrix would also be useful. With
\beq
\delta \ln G_k = t_{ka} \sigma_a^{-1} \delta Y_a,
\eeq
we could immediately infer the fractional change in fundamental couplings corresponding to a deviation of one abundance $Y_a$ away from the standard predicted value, while keeping the other abundances fixed. For example, a 3$\sigma$ change in \lise, {\em i.e.}\ $\delta Y_{\rm 7Li}/\sigma_{\rm 7Li} = 3$, would be correlated with a simultaneous fractional change $3 t_{ka}$ in the couplings $G_k$. Of course, if we have more couplings than observed abundances, the matrix $T$ with elements $t_{ka}$ is not uniquely defined. We may, however, restrict the number of independently varying fundamental couplings, either by keeping some couplings fixed or by assuming relations between the $G_k$, for example motivated by grand unification. If $S$ is reduced in this way to a $n \times n$ matrix one simply has
\beq
T = S^{-1}.
\eeq
As a simple example, we may discard the variation of the Yukawa couplings such that 
\[
\Delta \ln \hat{m} = \Delta \ln \delta_q = \Delta \ln m_e = \Delta \ln \vev{\phi}.
\]
For this case we show the matrices $S$ (transposed) and $T$ in Table~\ref{ST3by3}.
\begin{table}
\centering
\begin{tabular}{|c||c|c|c|}
\hline
$\sigma_a^{-1}\partial Y_a/\partial\ln G_k$
                                   &  D  & \hefo & \lise \\ \hline
\hline
$G_{\rm N}$                        &  6.1 &  9.9  & -6.5  \\ \hline
$\alpha$                           & 24   &  52  & -100   \\ \hline
$\vev{\phi}$                       & 100  & -125 & -560   \\ \hline
\end{tabular}
\hspace*{0.5cm}
\begin{tabular}{|c||c|c|c|}
\hline
$t_{ka}$ 
                                   &  D      & \hefo    & \lise    \\ \hline
\hline
$G_{\rm N}$                        & 0.24    & -0.015   &  0.048   \\ \hline
$\alpha$                           &-0.036   &  0.016   & -0.010   \\ \hline
$\vev{\phi}$                       & 0.0038  & -0.0027  & -0.00048 \\ \hline
\end{tabular}
\caption{Sensitivity matrix $S$ and its inverse $T$, given constant Yukawa couplings}\label{ST3by3}
\end{table}
A decrease of $6\sigma$ for \lise\ brings its predicted abundance from $4.5\times 10^{-10}$ to $1.5\times 10^{-10}$. With other abundances fixed, this could (in linear approximation) be realized by the variations
\beq
\Delta \ln(G_{\rm N}, \alpha, \vev{\phi}) = (-29\%, +6.1\%, +0.29\%).
\eeq
The required changes in $G_{\rm N}$ and $\alpha$ are relatively large: our linear treatment may still be valid here, but in general this should be checked, which we do explicitly for specific unified models in Section \ref{nonlinear}.
Alternatively, for a $5\sigma$ decrease in \lise\ from the standard predicted value and a $1\sigma$ increase in deuterium, bringing the predicted $Y_D$ to $3.0\times 10^{-5}$, we would require variations of fundamental parameters
\beq
\Delta \ln(G_{\rm N}, \alpha, \vev{\phi}) = (0, +1.4\%, +0.62\%),
\eeq
clearly well within the linear regime with respect to variation of fundamental parameters.

\section{Unified models}

It is of interest to consider unified scenarios where the variations of fundamental couplings satisfy relations that reduce the number of free parameters. In the simplest case every variation of a parameter $G_k$ is determined by a single underlying degree of freedom. The variations can then be written as a vector:
\beq
 \Delta \ln G_k = d_k \Delta \varphi
\eeq
where $\varphi$ is a (dimensionless) field which gives rise to the variation and $d_k$ are a set of numbers characterising a particular unified model. We then obtain
\beq
 \Delta \ln Y_a = (CF)_{ak} d_k \Delta \varphi,
\eeq
where we may also eliminate $\Delta \varphi$ in favour of the variation of some observable parameter. 

We will consider four simple possibilities. First, that the strength of gravitation varies, but all other scales and couplings of particle physics are unchanged. This corresponds to a violation of the strong equivalence principle, while the weak equivalence principle is preserved. Thus $d_k$ has a single nonzero entry corresponding to $G_{\rm N}$ and we find 
\beq
 \Delta \ln (Y_{\rm D}, Y_{3\rm He}, Y_{4\rm He}, Y_{6\rm Li}, Y_{7\rm Li}) \simeq 
 (0.94, 0.33, 0.36, 1.4, -0.72) \Delta \ln (G_{\rm N} \Lambda_c^2)
\eeq
restoring the implicit dependence on the QCD scale.

In the remaining examples, we consider a grand unified theory with unified coupling $\alpha_X$, broken at the scale $M_X$ to the Standard Model symmetry. The observable couplings of QCD and electromagnetism are then related to $\alpha_X$ via renormalization group (RG) flow and electroweak symmetry breaking. Thus one can find a relation between the variation of $\alpha$ and that of $\Lambda_c/M_X$, depending on the gauge group and matter content of the theory \cite{CalmetLangacker,DentF,WetterichProbing}.

We also need to specify the behaviour of the ratios of energy scales $M_X/M_{\rm P}$, $\vev{\phi}/M_{\rm P}$, $m_{e,q}/M_{\rm P}$, where $M_{\rm P}$ is the Planck mass proportional to $G_{\rm N}^{-1/2}$. In all unified scenarios we will take the Planck mass fixed relative to the unification scale, thus $\Delta (M_{\rm P}/M_X)=0$. For simplicity we take the Yukawa couplings to be constant, thus the electron and quark masses are proportional to $\vev{\phi}$. We also define an exponent $\gamma$ which relates the variation of $\vev{\phi}$ with respect to $M_X$ to the variation of $\Lambda_c/M_X$ as
\beq
 \frac{\vev{\phi}}{M_X} = \mbox{const.}\left(\frac{\Lambda_c}{M_X}\right)^{\gamma}.
\eeq

\subsection{Fixed ratio of weak and strong scales}
In the first unified scenario ``GUT1'' we take all low-energy 
mass scales of particle physics to be proportional to $\Lambda_c$, thus $\gamma=1$. 
Then using the non-supersymmetric GUT relations discussed in \cite{MSW,WetterichCrossover,WetterichProbing} we have
\bea
 \Delta \ln \alpha &=& \frac{22 \alpha}{7 \alpha_X} \Delta \ln \alpha_X \simeq 0.92 \Delta \ln \alpha_X, \\
 \Delta \ln \frac{\Lambda_c}{M_X} &=& \frac{\pi}{11\alpha} \Delta \ln \alpha \simeq 39 \Delta \ln \alpha
\eea
taking $\alpha_X\simeq 1/40$. These relations include the effects of varying (relative to $M_X$) charged particle masses, or ``thresholds'',  on the variation of gauge couplings.\footnote{In supersymmetric unified theories the effect of varying thresholds (charged particle masses) on the gauge couplings may be much larger \cite{Dentthresh}.} 
Then we have
\beq \label{dGGUT1}
 \Delta \ln(G_{\rm N},\alpha,\vev{\phi},m_e,\delta_q,\hat{m}) \simeq (78,1,0,0,0,0) \Delta \ln \alpha.
\eeq
In this case the variations of abundances are not subject to the theoretical uncertainty of varying $m_s/\Lambda_c$. 
We obtain 
\beq
 \Delta \ln (Y_{\rm D}, Y_{3\rm He}, Y_{4\rm He}, Y_{6\rm Li}, Y_{7\rm Li}) \simeq 
 (77, 27, 30, 120, -68) \Delta \ln \alpha
\eeq
where for definiteness we have taken all $f_i$ to unity in Eq.~(\ref{dBdmpi}). Note that when the variations are reexpressed in terms of $\Delta\ln G_{\rm N}$ as
\beq
 \Delta \ln (Y_{\rm D}, Y_{3\rm He}, Y_{4\rm He}, Y_{6\rm Li}, Y_{7\rm Li}) \simeq 
 (0.99, 0.34, 0.38, 1.5, -0.87) \Delta \ln G_{\rm N}
\eeq
the result is very similar to the first scenario where only $G_{\rm N}$ is varying. The scenarios only differ by a variation $\Delta \alpha/\alpha = (1/78) \Delta G_{\rm N}/G_{\rm N}$, hence we do not plot separately the first scenario of varying only $G_{\rm N}$.

\subsection{Fixed weak scale and varying strong scale}
In the second unified scenario ``GUT2'' we consider that the Higgs v.e.v.\ and elementary fermion masses are all proportional to the unification scale, thus $\Delta(M_X/M_{\rm P},\vev{\phi}/M_{\rm P},m_{e,q}/M_{\rm P})=0$, or equivalently $\gamma=0$. Then, after converting to QCD units, the mass scales $M_{\rm P}$, $\vev{\phi}$ and $m_{e,q}$ will vary inversely to $\Lambda_c/M_X$. We find \cite{MSW}
\bea
 \Delta \ln \alpha(M_W) &=& \frac{8 \alpha}{3 \alpha_X} \Delta \ln \alpha_X, \\
 \Delta \ln \frac{\Lambda_c}{M_X} &=& \Delta \ln \frac{\Lambda_c}{\vev{\phi}}
 = \frac{\pi}{12\alpha} \Delta \ln \alpha(M_W).
\eea
Due to the effect of the three light quarks whose effect on the running of $\alpha(\mu)$ is cut off at $\mu \sim \Lambda_c$, the variation of the fine structure constant is as follows
\beq
 \frac{1}{\alpha} \Delta \ln \alpha = \frac{1}{\alpha(M_W)} \Delta \ln \alpha(M_W) \left (1+\frac{1}{18}\sum_i \tilde{Q}_i^2 \right)
\eeq
where $i$ runs over three colors of $u$, $d$, and $s$ quark, thus $\sum_i \tilde{Q}_i^2 =2$. Then the variations of fundamental couplings are related as
\beq 
 \Delta \ln \frac{\Lambda_c}{M_X} = \Delta \ln \frac{\Lambda_c}{\vev{\phi}}
 = \frac{3\pi}{40\alpha} \Delta \ln \alpha \simeq 32.3 \Delta \ln \alpha
\eeq
and we have
\beq \label{dGGUT2}
 \Delta \ln(G_{\rm N},\alpha,\vev{\phi},m_e,\delta_q,\hat{m}) \simeq (64.5,1,-32.3,-32.3,-32.3,-32.3) \Delta \ln \alpha. 
\eeq
We then obtain variations of abundances
\beq
 \Delta \ln (Y_{\rm D}, Y_{3\rm He}, Y_{4\rm He}, Y_{6\rm Li}, Y_{7\rm Li}) \simeq 
 (-450, -130, 170, 380, 1960) \Delta \ln \alpha
\eeq
Note that this model is subject to additional uncertainty due to the variation in the strange quark mass relative to $\Lambda_c$; however it seems unlikely that this variation would produce significant cancellations.

\subsection{Varying the weak scale faster than the strong scale}
In the third unified scenario ``GUT3'' we consider the case when the Higgs v.e.v.\ and fermion masses vary {\em more}\/ rapidly (with respect to the unification scale) than the QCD scale $\Lambda_c$ does: thus $\gamma > 1$. We take $\gamma = 1.5$ and find that the variations of fundamental couplings are related as
\beq \label{dGGUT3}
 \Delta \ln(G_{\rm N},\alpha,\vev{\phi},m_e,\delta_q,\hat{m}) \simeq (87,1,21.5,21.5,21.5,21.5) \Delta \ln \alpha. 
\eeq
The variations of abundances are then
\beq
 \Delta \ln (Y_{\rm D}, Y_{3\rm He}, Y_{4\rm He}, Y_{6\rm Li}, Y_{7\rm Li}) \simeq 
 (430,130,-65,-60,-1420) \Delta \ln \alpha.
\eeq

\subsection{Results}
\begin{figure} 
\begin{center}\vspace*{-2cm}
\includegraphics[width=8.3cm]{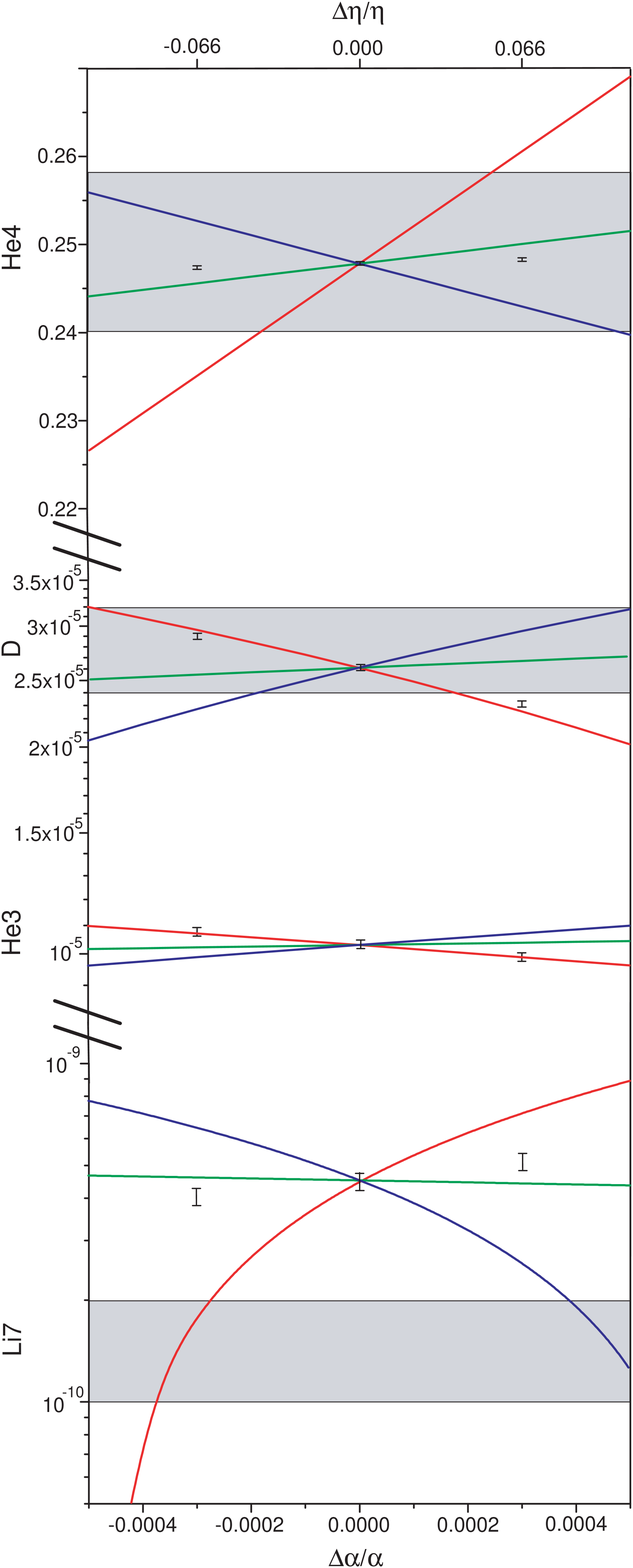}
\end{center}
\caption{Variation of primordial abundances with $\alpha$ in three GUT scenarios. Green lines (with smallest variations of abundances) show the first ``GUT1''; red lines (with large positive slope of $Y_{\rm 7Li}$) the second ``GUT2''; and blue lines (with large negative slope of $Y_{\rm 7Li}$) the third ``GUT3''. Highlighted regions give the observational 1$\sigma$ limits. Error bars indicate the standard BBN abundances with theoretical 1$\sigma$ error \cite{Serpico04}, for three different values of $\eta$ about the WMAP central value, as indicated on the upper horizonal axis.}\label{FigVaryGUT}
\end{figure}

In Fig.~\ref{FigVaryGUT} we show the abundance variations given by the three GUT models, as a function of the variation of $\alpha$. Note that we plot only the linear dependence of abundances on $\alpha$, therefore if $\Delta \ln Y_a$ becomes larger than 1 (as in the case of \lise) the results may be affected by higher order terms. We also show the 1$\sigma$ observational bounds as highlighted regions. Also included in the plot is the effect on the standard BBN predictions of varying the baryon-to-photon ratio $\eta$ over the 2$\sigma$ range allowed by WMAP 3 year data, $5.7\leq 10^{10}\eta \leq 6.5$. 

It can be seen that in the ``GUT2'' scenario a reduction of $\alpha$ by about $0.025 \%$ ({\it i.e.}\ a fractional variation of $-2.5\times 10^{-4}$) would bring theory and observation into agreement within 2$\sigma$ bounds, while remaining in the linear regime. Conversely, in the ``GUT3'' model an increase of $\alpha$ by about $0.04 \%$, {\it i.e.}\ $\Delta \ln \alpha = 4\times 10^{-4}$, brings theory and observation into agreement within 1$\sigma$ bounds. 
Considering the variations of fundamental parameters in the three scenarios Eqns.~(\ref{dGGUT1},\ref{dGGUT2},\ref{dGGUT3}), the behaviour of the weak scale $\vev{\phi}$ and fermion masses is decisive for the variation of abundances.

\subsection{Nonlinear variation of abundances in GUT scenarios} \label{nonlinear}

The unified models discussed in the previous section suggest that it is possible, and may even be natural, to obtain a large negative variation in the \lise\ abundance, and considerably smaller variations in other measurable abundances: positive in the case of deuterium and negative for \hefo. Agreement between theory and data in all three abundances could then be possible for a narrow range of values in the variation of fundamental parameters, and such scenarios could be tested by more accurate abundance measurements. However, the required fractional variation in \lise\ is so large (a factor two or more in $Y_{7\rm Li}$) that a linear analysis using matrix multiplication may be inaccurate.  

We may improve the analysis in specific cases by including the nonlinear relations between nuclear parameters and abundances. This is implemented simply by running the numerical integration code with the appropriate values of nuclear parameters, where the dependence on nuclear parameters was detailed in Section \ref{nucparams}. This method would be impractical to investigating the full parameter space: nuclear parameters span a 13-dimensional space (or 12-dimensional if \lisi\ is neglected). It is only practicable if the dimensionality of the parameter space is reduced by applying unification relations. In principle we could also attempt to estimate the nonlinear dependence of nuclear parameters $X_i$ on fundamental parameters $G_k$, but this involves additional theoretical uncertainty. For the unified models considered here, the fractional variations in $X_i$ remain small, well below $0.1$. A linear approximation for the relation between nuclear and fundamental parameters is therefore appropriate. 
The main nonlinear effects enter at the level of nuclear reactions. 

The nuclear parameters affecting most the large variation in \lise\ abundance are mainly the deuterium and \bese\ binding energies, with the \heth\ and \hefo\ binding energies playing a smaller r{\^ o}le. A decrease of $B_{\rm D}$ causes BBN to happen later, which means that the nucleon density is lower and reaction rates smaller. The abundances of $A>4$ elements are rate-limited and thus decrease with decreasing $B_{\rm D}$. This accounts for about two-thirds of the change in $Y_{7\rm Li}$. In addition, the cross-section of the \heth$(\alpha,\gamma)$\bese\ reaction depends strongly on the $Q$-value, hence on the \bese\ binding energy. Both these effects are computationally under control, therefore we believe that the specific nonlinear dependence in the scenarios we consider is well estimated within our code.

\begin{figure} 
\begin{center}\vspace*{-2cm}
\includegraphics[width=8.3cm]{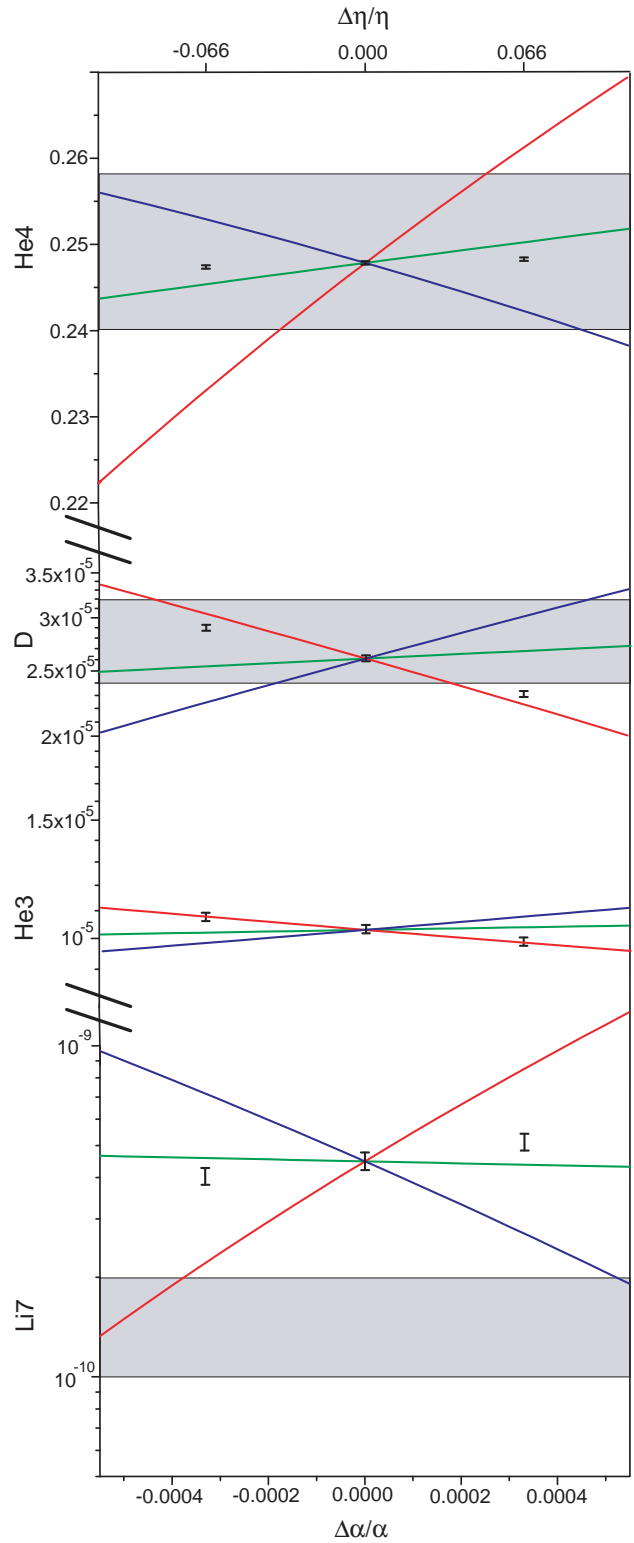}
\end{center}
\caption{Variation of primordial abundances with $\alpha$ in three GUT scenarios including nonlinear effects. Labels as in Fig.~1.}\label{FigVaryGUTnonlin}
\end{figure} 
We show in Figure~\ref{FigVaryGUTnonlin} the primordial abundances including nonlinear effects, {\em i.e.}\ without using a linear approximation for the relation between $Y_a$ and $X_i$. For our three GUT models we find a slightly different behaviour of the \lise\ abundance, which now has an approximately power-law dependence on variation of $\alpha$. It is only slightly more difficult to bring the present observational abundances into agreement with standard BBN and the WMAP determination of $\eta$ ; still, if we allow a variation of $0.00045 \lesssim \Delta \ln \alpha \lesssim 0.0005$ in the ``GUT3'' model, the predicted abundances are all very close to the 1$\sigma$ allowed regions.

\section{Conclusions}

We have developed a systematic method to relate cosmological variations in the underlying parameters of particle physics to the primordial isotope abundances produced by BBN. The main advantage of the method is that we are able to vary every parameter independently, both at the level of fundamental Standard Model parameters and of nuclear physics parameters, thus we are not dependent on any particular theoretical model which enforces particular relations between the variations. 

The method proceeds by defining two linear response matrices. The first, $C$, encodes the change in predicted abundances produced by small variations away from the current values of nuclear physics parameters which enter the BBN integration code. These parameters (in units where the nonperturbative QCD scale is constant) comprise the gravitational constant, fine structure constant, neutron lifetime, electron, proton and neutron masses, and binding energies of $A\leq7$ nuclei. The dependences of nuclear reaction rates on these parameters are also implemented insofar as they are calculated within some effective theory. One notable result is that the \lise\ abundance depends heavily on the binding energies of \heth, \hefo\ and \bese.

We also investigated possible further effects of variations in nuclear reaction rates on predicted abundances by varying each rate ({\it i.\,e.}\ thermal integrated cross-section $\vev{\sigma v}$) separately by a temperature-independent factor. We find that the \hefo\ abundance is insensitive to nuclear rates, and only eight reactions could lead to significant variation of the D, \heth\ or \lise\ abundances. Also in many cases the dependences on $\vev{\sigma v}$ are small and probably subleading compared to other known effects of varying nuclear parameters.

The second matrix $F$ relates variations in nuclear parameters to the fundamental parameters of particle physics, comprising the gravitational constant, fine structure constant, Higgs vacuum expectation value, electron mass, and the light (up and down) quark masses. At this point theoretical uncertainty enters into the relation between quark masses and nuclear binding energies. We parameterise the dependence of binding energies on the pion mass (and hence on light quark masses) by the deuteron binding, which has been treated by a systematic expansion in effective field theory. 

The resulting fundamental response matrix $R=CF$ allows us, first, to bound the variations of the six fundamental couplings individually, some bounds being at the percent level. We then define a ``sensitivity matrix'' that shows which observational data give the best determination of each fundamental parameter, and demonstrate the use of the inverse sensitivity matrix to find the variations in fundamental couplings required by any given change in primordial abundances. 

We can also bound correlated variations affecting many couplings at once: we consider three simple scenarios motivated by grand unification of gauge couplings. Of these, one allows us to fit observed D, \hefo\ and \lise\ abundances within $2\sigma$ bounds, given a variation $\Delta\alpha/\alpha = -2\times 10^{-4}$ away from the present value; another fits these observational abundances within 1$\sigma$ bounds, given a variation $\Delta\alpha/\alpha = 4\times 10^{-4}$.

Progress in the field requires both observational and theoretical improvements. Both statistical and systematic errors in abundance measurements could be improved, for example observations to better determine the nature of systems where \hefo\ is measured \cite{Steigman05}, or stellar modelling to test possible solutions of the \lise\ problem. On the theoretical side the relation between quark masses and nuclear physics remains unclear beyond the level of the two-nucleon system: the largest uncertainty in our BBN bounds arises from the poorly known dependence of the binding energies on the fundamental couplings.

BBN is already the most powerful probe of fundamental ``constants'' in the early Universe, and precision bounds may well be obtained, given continued efforts in observation and theory, to rule out or confirm the presence of a cosmological variation.

\subsection*{Acknowledgements}
We acknowledge useful discussions with P.~Descouvemont, R.~Cyburt, N.~Nunes, J.~Kneller and J.-W.~Chen. T.\,D. is supported by the {\em Impuls- and Vernetzungsfond der Helmholtz-Gesellschaft}.

\appendix

\section{Implementation of the BBN code and reaction rates}\label{AppendixA}

\subsection{Units, changes to the code and general remarks}\label{unitscode}
The code is written in terms of units MeV and $10^9$ Kelvin. We have to make a choice of how to define these units, in a context where ratios of dimensionful quantities may be varying. We choose to define units such that the QCD scale $\Lambda_c$ is constant: thus $\Lambda_c$ is always the same number of MeV, and this energy scale always corresponds to the same temperature in Kelvin. This simplifies the treatment of nuclear physics and QCD; we discuss below in Section \ref{nucscaling} what must be done to translate to other units, or to a unit-free formulation. All numerical constants and conversion factors in the code have been replaced with variables which are given their values globally, allowing a consistent variation to be implemented.

For the $n \leftrightarrow p$ reaction rates, instead of analytic approximations \cite{Kawano92} we implement exact formulae \cite{Scherrer83} incorporating also zero-temperature and thermal Coulomb corrections \cite{LopezTurner99}. The default baryon density is taken from the WMAP3 determination $\eta = 6.1\times 10^{-10}$ \cite{Spergel03}. Note that this determination of $\eta$ assumes that the values of couplings and mass scales are the same as today. Since we keep $\eta$ as a parameter, the possible effect of varying couplings at the time of CMB decoupling could be incorporated into our analysis.

Current measurements of the neutron lifetime are strongly inconsistent, the most recent being $878.5\pm 0.7\,$s \cite{Serebrov07} compared to the 2006 PDG world average $885.7\pm0.8\,$s \cite{PDG06}. Since this discrepancy is less than 1\% and no abundance depends very strongly on $\tau_n$, we take the lifetime to be the PDG value for the purpose of calculating the leading dependence of abundances. 

Starting with the Kawano 1992 code \cite{Kawano92}, we replaced all (rounded) numerical constants in the code by functions of natural ``constants'', which are globally defined as variables. Thus, variations of constants are treated consistently within the code.
Thanks to increased computational power
we can remove all significant sources of numerical inaccuracy: these were for example 32 bit internal precision of floating point numbers, simplified integration routines, large time steps 
to name just a few. Better numerical accuracy allows us to study the behaviour of the abundances under very small changes of ``constants'', which is essential to derive the linear 
variation about the standard prediction, and thus the derivatives of abundances,
without numerical ambiguity.

\subsection{Fits for charged particle reactions}
Reaction rates for charged particles (with atomic numbers $Z_{1,2}$ in the initial state) arise from a thermal average of a cross-section 
which in the absence of resonances is the product of the Gamow factor and an ``S-factor'': 
\beq
\sigma (E) = S(E) \frac{e^{-2\pi\tilde{\eta}}}{E}
\eeq
where $\tilde{\eta} \equiv \alpha Z_1Z_2\sqrt{\mu/2E}$ and $\mu$ is the reduced mass. The S-factor may be expanded in a Maclaurin series to quadratic order in energy, which is usually sufficient to account for any smoothly-varying dependence. However some cross-sections are fit with an additional exponential term $\tilde{S}(0)e^{-\beta E}$ \cite{FCZ2}. In addition, non-resonant terms may be multiplied by a cutoff factor $f_{\rm cut}=e^{-(T/T_{\rm cut})^2}$, where $T_{\rm cut}$ has been argued to be proportional to $\alpha^{-1}$ \cite{BergstromIR,FCZ2}.

Where the cross-section as a function of energy shows one or more resonances, they contribute to the thermal averaged rate as
\beq
\vev{\sigma v}_{\rm res} = g(T)e^{-\bar{E}/T}
\eeq
where $g(T)$ and $\bar{E}$ are fitting parameters corresponding to the shape and position of the resonance. Usually a power-law is taken for $g(T)$, thus $g(T)=c T^p$. In principle one should consider the variation of the resonance parameters if this term is significant. But since the major contribution to the resonance energy $\bar{E}$ probably arises from $\Lambda_c$, the invariant scale of strong interactions, which we take as our (non-varying) unit, it seems a reasonable first guess to keep the resonance parameters fixed.

In order to treat the $\alpha$-dependence of reaction rates consistently, so far as it can be calculated, we must implement it at the level of the cross-section $\sigma$. Hence the NACRE formulae \cite{NACRE} fitted at the level of the thermal averaged cross-sections $\vev{\sigma v}$ are not suitable. Instead we used the functional forms of rates from \cite{SKM93, BergstromIR}, which correspond to a definite S-factor expansion for $\sigma$, in addition to resonant terms. The free parameters, primarily the expansion coefficients of $S(E)$ and resonance parameters, were then fit, in most cases to reproduce the NETGEN rates \cite{NETGEN} as closely as possible. We also checked that the resulting cross-sections are consistent with experimental data. In the case of $d(\alpha,\gamma)$\lisi\ we found a set of parameters which seems to fit the experimental cross-section at low energies \cite{Kiener6Li} better than the NACRE fit. But note that this cross-section is not measured directly, rather it is derived from experimental data under various assumptions, which should be more carefully investigated \cite{Typeltalk}.

Replacing the NETGEN rates in the code with our fitted rates, we obtain abundances which are changed as follows: $Y_{\rm D}$ differs by $-0.3\%$, $Y_{3\rm He}$ by $+0.9\%$, $Y_{4\rm He}$ by less than $0.1\%$, $Y_{7\rm Li}$ by $+3\%$. Hence we do not consider this refitting as significant, except in the case of the $d(\alpha,\gamma)$\lisi\ reaction. Depending on whether this reaction was fit to NETGEN, or to the cross-section values of \cite{Kiener6Li}, we found a \lisi\ abundance larger by a factor of 1.02, or 3.3, respectively. Given the unclear observational status of \lisi\ this discrepancy is not currently worth pursuing.

\subsection{Scaling of dimensionful nuclear parameters} \label{nucscaling}
The variation of nuclear binding energies and reactions involve (in most cases) only two mass scales: the QCD invariant scale $\Lambda_c$, and the average light quark mass $\hat{m}$, which indicates the departure from the chiral limit $m_q \rightarrow 0$. In the chiral limit the dependence of binding energies and strong interaction cross-sections becomes extremely simple: all dimensionful parameters are simply proportional to a power of $\Lambda_c$. Switching on the quark masses, one obtains a finite range for pion-mediated interactions, which may greatly affect static and dynamical properties of nuclei. Also, the masses of all hadrons are affected at some order in chiral perturbation theory \cite{GasserLeutwyler}. However, if both $\Lambda_c$ and $m_q$ are varied by a common factor, while all dimensionless couplings are held constant, dimensionful quantities involving strong interactions (and to a good approximation electromagnetic interactions) scale with some power of this common factor. Such a variation of dimensionful parameters is equivalent to a redefinition or variation of the units of mass or energy.

Since we fix our unit of mass to be a constant times $\Lambda_c$, the correct behaviour of dimensionful quantities associated with QCD or the strong nuclear force is automatic. So long as quark masses are proportional to $\Lambda_c$, such quantities are unchanged. The effects of other dimensionful parameters $\vev{\phi}$ and $G_{\rm N}$ are then transparent.

If, however, $\Lambda_c$ is formally allowed to vary, the scaling properties of QCD, and hence of the strong nuclear force, provide a simple check on the dependence of physical quantities on dimensionful parameters. For example, we will write the dependence of the deuteron binding energy $B_{\rm D}$ on the pion mass as
\beq
 \Delta \ln B_{\rm D} = r \Delta \ln m_\pi
\eeq
in units where $\Lambda_c$ is constant. We may thus rewrite variations of dimensionful quantities as, for instance, $\Delta \ln (B_{\rm D}/\Lambda_c)$ and obtain
\beq
 \Delta \ln B_{\rm D} = r \Delta \ln m_\pi + (1-r) \Delta \ln \Lambda_c,
\eeq
allowing us to compare results obtained with different choices of unit. 
For this type of equation relating fractional variations of dimensionful quantities, there is a simple check on whether it behaves correctly under redefinition of units. Each term is a numerical coefficient $r_i$ times the fractional variation $\Delta \ln Q_i$ of a quantity $Q_i$ with mass-energy dimension $D_i$: then the sum of $r_iD_i$ on each side of the equation must match.

\section{Observational situation and uncertainties}\label{AppendixB}

One of the biggest success of standard BBN is the matching of theoretically predicted and observed primordial abundances for major elements. For a review of the theoretical and observational status and obstacles see \cite{Steigman05}. 
The highest precision measurement is that of the \hefo\ abundance (conventionally written $Y_P$); however 
the actual precision of the determinations and possible systematic errors are currently debated \cite{Steigman05}. The situation is exemplified by recent contradictory observational determinations 
\cite{Izotov07+Peimbert}. It was argued in \cite{OliveSkillman04} that given a range of systematic effects the observational data indicated a primordial abundance of
\begin{equation}
Y_P = 0.249 \pm 0.009.
\end{equation}
which we take to be a 1$\sigma$ range. However, given the probable dominance of systematic effects, instead of using 2-$\sigma$ bounds to determine the range of allowed variations, we rather use the ``conservative allowable range'' of $Y_P$ given in \cite{OliveSkillman04} as
\beq
0.232 \le Y_P \le 0.258.
\eeq

The determination of the primordial deuterium abundance follows from a small number of observed systems. Recent determinations \cite{OMeara06,Kirkman03} yield a value of 
\begin{equation}
\mbox{D/H} = (2.8 \pm 0.4) \times 10^{-5}
\end{equation}
where the large scatter between values determined from different systems should be noted.

Determinations of the \heth\ abundance typically have a large scatter. Considering also the complex post-BBN development of this isotope, \heth\ abundance determinations cannot be considered as good tracers for the primordial abundance \cite{Vangioni-Flam02}.

Observations of the \lise\ abundance show a plateau at old, metal-poor halo stars, suggesting that the plateau value is closely related to the primordial one. The most recent determinations of the abundance are quite small: \lise/H = $(1.3 \pm 0.3) \cdot 10^{-10}$ \cite{Bonifacio06} (see also \cite{Asplund05}). It has been suggested that there are unresolved systematic errors relating to the effective temperature of the stars \cite{Charbonnel05} which may imply a value as large as \lise/H = $(1.64 \pm 0.3) \cdot 10^{-10}$.
To account for this possible systematic, we adopt a value
\begin{equation}
\mbox{\lise/H} = (1.5 \pm 0.5) \times 10^{-10}.
\end{equation}
Thus the observed \lise\ abundance is about a factor of 3 smaller than the standard theoretical prediction. 

A possible detection of \lisi\ was discussed in \cite{Asplund05}, though not at high statistical significance. 
If the detection is correct, the \lisi\ abundance is about a factor 1000 larger than the SBBN prediction. Given the unclear observational status and post-BBN history of the isotope, we do not include \lisi\ in the final analysis.

Theoretically predicted primordial abundances also come with an error, mainly due to cross-section uncertainties. Our numerical procedures do not provide error estimates, so we adopt the $1\sigma$ ranges from \cite{Serpico04}, using a baryon density $\Omega_b h^2=0.0224$ \cite{Manganotalk}:
\bea
\mbox{D/H}     &=& (2.61 \pm 0.04) \times 10^{-5} \nonumber \\
\mbox{\heth/H} &=& (1.03 \pm 0.03) \times 10^{-5} \nonumber \\
Y_P            &=& 0.2478 \pm 0.0002 \nonumber \\
\mbox{\lise/H} &=& (4.5  \pm 0.4 ) \times 10^{-10}.
\eea

\end{document}